\documentclass[conference]{IEEEtran}
\IEEEoverridecommandlockouts

\usepackage{cite}
\usepackage{amsmath,amssymb,amsfonts}
\usepackage{graphicx}
\usepackage{textcomp}
\usepackage{xcolor}
\def\BibTeX{{\rm B\kern-.05em{\sc i\kern-.025em b}\kern-.08em
    T\kern-.1667em\lower.7ex\hbox{E}\kern-.125emX}}

\usepackage{booktabs}
\usepackage{multirow}
\usepackage{xcolor}
\definecolor{deltared}{RGB}{190, 80, 70}
\definecolor{deltagreen}{RGB}{60, 140, 80}

\newcommand{\dn}[1]{{\scriptsize\color{deltared}\,({#1})}}    
\newcommand{\up}[1]{{\scriptsize\color{deltagreen}\,(+{#1})}} 

\usepackage{url}
\usepackage{algorithm}
\usepackage{algpseudocode}

\begin{document}


\title{Vision Token Manipulation Attacks on Cloud-Edge Inference of Large Vision-Language Models
}

\author{
\IEEEauthorblockN{Zikai Zhang$^{1}$, Rui Hu$^{1}$, Olivera Kotevska$^{2}$, Jiahao Xu$^{1}$}
\IEEEauthorblockA{$^{1}$Department of Computer Science and Engineering, University of Nevada, Reno, Reno, USA\\
$^{2}$Oak Ridge National Laboratory, Oak Ridge, USA\\
zikaiz@unr.edu, ruihu@unr.edu, kotevskao@ornl.gov, jiahaox@unr.edu}
}

\maketitle

\begin{abstract}
Cloud-edge Large Vision-Language Model (LVLM) inference enables efficient deployment by splitting computation between edge devices and cloud servers. In this process, intermediate vision tokens are transmitted from the edge to the cloud over a communication link, thereby exposing a new attack surface. We study vision token manipulation attack (VTM-Attack) under a black-box man-in-the-middle setting, where an adversary intercepts and manipulates a subset of transmitted vision tokens under a budget constraint. We propose four na\"ive attack strategies and an optimization-based token selection method. Experiments on 6 state-of-the-art LVLMs (3B-72B) across 4 benchmarks show that manipulating only 10\% of vision tokens can reduce accuracy by up to 88.31\%. These results reveal a critical vulnerability in cloud-edge LVLM inference.
\end{abstract}


\section{Introduction}

Large Vision-Language Models (LVLMs)~\cite{Qwen2.5-VL,wang2025internvl3} have recently achieved remarkable progress on a wide range of vision-language tasks, including optical character recognition (OCR)~\cite{liu2024ocrbench}, visual question answering (VQA)~\cite{liu2024mmbench}, instruction following~\cite{liu2023hallusionbench}, and mathematical reasoning~\cite{lu2023mathvista}. By coupling a vision encoder with a Large Language Model (LLM), LVLMs can transform visual inputs into semantic representations and perform vision-language interaction.

Due to their strong capabilities and broad applicability, LVLMs are increasingly being deployed in practical cloud-edge environments~\cite{li2025distributed,qian2025edgevlm}. In such architectures, resource-constrained edge devices perform lightweight visual encoding and generate vision tokens, which are then transmitted to cloud servers for further vision-language interactions. Computationally intensive components, including the text encoder and the LLM backbone, are deployed in the cloud. This design substantially reduces the computational cost on resource-constrained edge devices without sacrificing performances.

However, this cloud-edge architecture introduces a new attack surface absent in centralized deployments: the communication link that transmits vision tokens from the edge to the cloud. This link is particularly vulnerable to man-in-the-middle adversaries~\cite{conti2016survey}, who can intercept and manipulate messages (e.g., vision tokens) transmitted over communication channels. Notably, vision tokens form a compact, high-dimensional representation that directly encodes semantic content, making them an effective target for manipulation at the representation level.
For example, an adversary can manipulate the ordering of transmitted vision tokens to disrupt the structural consistency of the visual representation, thereby misleading the LVLM’s understanding of spatial relationships. Meanwhile, the attacker can alter a small portion of tokens that encode critical semantic cues, effectively shifting the representation of the original visual input.
These properties highlight the potential risks of the communication link in cloud-edge LVLM inference systems.


Existing studies on LVLM security~\cite{goodfellow2014explaining,li2025backdoorvlm} mainly focus on attacks on visual inputs. For instance, adversarial perturbations~\cite{goodfellow2014explaining} aim to mislead the model by modifying raw images through carefully crafted noise, while backdoor attacks~\cite{li2025backdoorvlm} embed specific visual patterns to induce targeted model behaviors. These approaches assume direct access to raw visual data and operate entirely in the input space before encoding.
Meanwhile, existing works study security risks in cloud-edge inference for LLMs, investigating adversarial~\cite{fan2023robustness} and inference~\cite{luo2025prompt} attacks on intermediate representations of text prompts. These approaches reveal vulnerabilities in transmitted representations but focus on textual modalities and overlook vision-language settings.
As a result, the security implications of manipulating transmitted vision tokens in cloud-edge LVLM inference systems remain largely unexplored.

In this paper, we study \textbf{V}ision \textbf{T}oken \textbf{M}anipulation Attack (\textbf{VTM-Attack}) on cloud-edge LVLM inference systems. We consider a man-in-the-middle adversary that intercepts transmitted vision tokens and replaces them with manipulated ones before forwarding them to the cloud. 
To ensure practical feasibility, the attack is subject to constraints on the properties of the manipulated vision tokens.
%
%
Based on this attack model, we propose a family of VTM-Attacks on transmitted vision tokens at the cloud-edge interface. We first design structure- and value-based manipulation strategies, including permutation, masking, Gaussian perturbation, and sign flip. We then develop an optimization-based token selection strategy based on self-attention disruption and norm-aware regularization losses.

To evaluate the effectiveness of VTM-Attacks, we conduct extensive experiments on six state-of-the-art LVLMs from the Qwen2.5-VL-Instruct~\cite{Qwen2.5-VL} and InternVL3.5~\cite{wang2025internvl3} families across four benchmarks, covering OCR~\cite{liu2024ocrbench}, VQA~\cite{liu2024mmbench}, instruction following~\cite{liu2023hallusionbench}, and mathematical reasoning~\cite{lu2023mathvista} tasks. Results reveal a critical vulnerability: under sign flip attacks, model accuracy can drop drastically (e.g., Qwen2.5-VL-72B from 88.39\% to 0.08\% on MMBench). The proposed optimization-based token selection further amplifies attack effectiveness, demonstrating significant performance degradation across models and benchmarks.

Our main contributions are summarized as follows:
\begin{itemize}
    \item We identify the cloud-edge vision token transmission interface as a new and practical attack surface and formulate a man-in-the-middle attack model for vision token manipulation.
    \item We propose VTM-Attacks, a family of token-level attacks comprising structure- and value-based manipulation strategies, including permutation, masking, Gaussian perturbation, and sign flip.
    \item We design an optimization-based token selection strategy based on self-attention disruption and norm-aware regularization losses, which significantly improves attack effectiveness by identifying these vulnerable vision tokens.
    \item We conduct extensive evaluations on six state-of-the-art LVLMs (3B-72B) across four benchmarks, demonstrating that cloud-edge LVLM inference is highly vulnerable to VTM-Attacks with even limited manipulation budgets.
\end{itemize}

The rest of this paper is organized as follows. In Section~\ref{sec:system}, we present the cloud-edge LVLM inference system model. In Section~\ref{sec:vtm-attack}, we describe the attack model, we also introduce four na\"ive VTM-Attack strategies and optimization-based token selection method. In Section~\ref{sec:experiments}, we present the experimental setup and empirical results. Finally, we conclude the paper in Section~\ref{sec:conclusion}.


\section{System Model}\label{sec:system}
\begin{figure}
    \centering
    \includegraphics[width=0.85\linewidth]{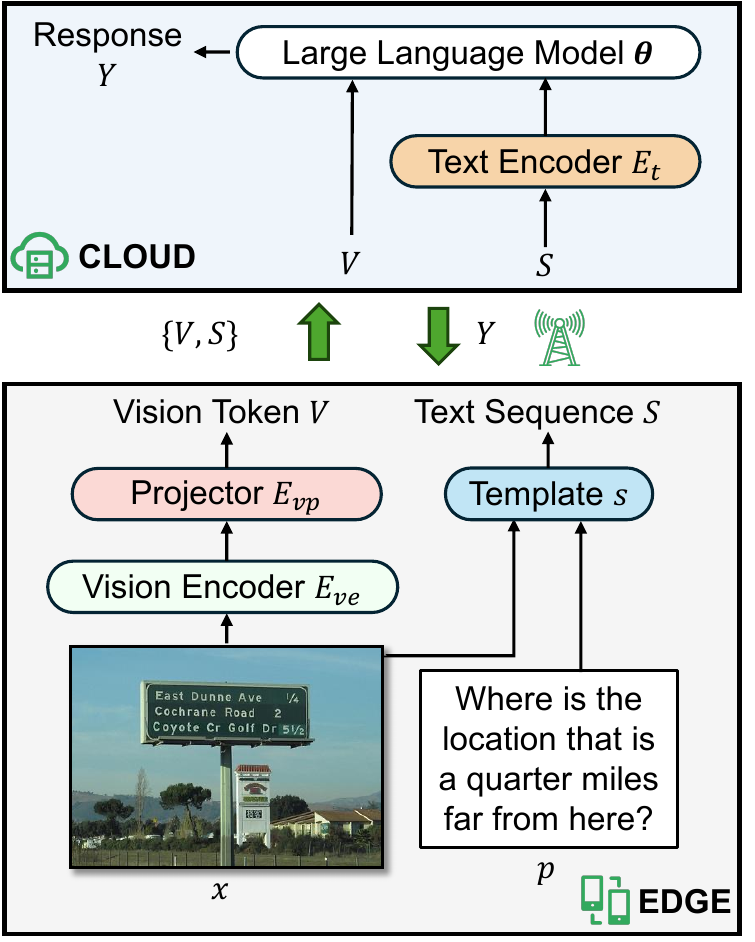}
    \caption{Overall framework of cloud-edge LVLM inference system.}
    \label{fig:system}
\end{figure}

We consider a cloud-edge LVLM inference system~\cite{li2025distributed,qian2025edgevlm} consisting of an edge device and a cloud server.
The system offloads computationally intensive model components to the cloud while keeping lightweight computation at the edge. Specifically, as shown in Fig.~\ref{fig:system}, the LVLM is partitioned into two modules: the edge device hosts the vision encoder and projector for visual feature extraction, while the cloud server hosts the text encoder layer and the LLM backbone for vision-language interaction and autoregressive response generation.

\noindent\textbf{Edge Module.}
Given a visual input $x$ and a text prompt $p$, the edge device extracts visual features via a vision encoder $E_{ve}$, followed by a projector $E_{vp}$. We denote the combined mapping as $E_v$, which produces vision tokens ${V} = E_v(x) = [v_1,\dots,v_{L_v}]^\top  \in \mathbb{R}^{L_v\times d}$. Here, $L_v$ denotes the number of vision tokens determined by the LVLM pre-processing pipeline, and $d$ is the vision embedding dimension.
Concurrently, the text prompt is incorporated into a vision-language template $s$ that interleaves textual content with visual placeholders, forming a sequence $S$ comprising text elements and visual placeholder elements. 
%
The edge transmits a composite signal to the cloud, defined as $\mathcal{T} := \{{V}, {S}\}$. 

\noindent\textbf{Cloud Module.}
Upon receiving $\mathcal{T}$, the cloud server tokenizes the sequence $S$ to obtain text tokens $K$ and then mapped into the embedding space via the text encoder layer $E_t$, yielding text embeddings $P=E_t(K)$. These embeddings are subsequently fused with the visual tokens ${V}$ at positions specified by the placeholder tokens.


The resulting vision-language representations are processed by the LLM parameterized by $\boldsymbol{\theta}$ for autoregressive decoding. Specifically, at the $t$-th decoding step, the next token $y_t$ is sampled from $y_t \sim \mathcal{P}(y_t | V, P, Y_{<t}; \boldsymbol{\theta})$, where ${Y}_{<t} = (y_1, \dots, y_{t-1})$ denotes the sequence of previously generated tokens. The decoding proceeds sequentially until a predefined termination condition is met.

\section{Vision Token Manipulation Attacks}
\label{sec:vtm-attack}

We present VTM-Attacks, a family of attacks on cloud-edge LVLM inference that target the communication link by manipulating the vision tokens.




\subsection{Attack Model}\label{subsec:attackmodel}

We consider a man-in-the-middle adversary~\cite{conti2016survey,piggott2023net} operating on the communication link between cloud and edge, which intercepts the transmitted signal $\mathcal{T}$. To maintain stealthiness, the adversary focuses exclusively on vision token manipulation, as text manipulation is more easily detectable~\cite{liu2024formalizing}. Specifically, it replaces $V$ with a manipulated version $\hat{V}$
before forwarding it to the cloud. 

\noindent\textbf{Attacker’s Goal and Background Knowledge:} The attacker aims to alter the model’s generation behavior in an untargeted manner without disrupting the normal inference pipeline. We assume a black-box setting in which the adversary has no access to the model on either the edge or the cloud. 

\noindent\textbf{Attacker’s Capability and Constraints:} 
The adversary modifies the vision tokens $V$ to produce $\hat{V}$ under the following constraints:
(i)~\emph{Dimensional consistency}: $\hat{V}$ must preserve the original shape $L_v \times d$ of $V$ to remain compatible with the vision-language processing pipeline; otherwise, the inference process fails and raises anomalies immediately; and (ii)~\emph{Manipulation budget}: the adversary can only modify a subset $\mathcal{I} \subseteq \{1, \dots, L_v\}$ of vision tokens, where $\mathcal{I}$ denotes the index set of manipulated vision tokens, a larger $|\mathcal{I}|$ leads to reduced stealthiness. 

\noindent\textbf{Attack Formulation:} Formally, the adversary seeks $\hat{V}$ such that
\begin{equation}\label{eq:overall_obj}
\begin{aligned}
&\mathcal{P}(y_t \mid \hat{V}, P, Y_{<t}; \boldsymbol{\theta}) \neq
\mathcal{P}(y_t \mid V, P, Y_{<t}; \boldsymbol{\theta}) \\
&\text{s.t. } \exists\;\mathcal{I} \subseteq \{1, \dots, L_v\}, \; |\mathcal{I}| \leq \rho L_v, \\
&\quad\;\; \hat{v}_i = v_i, \;\; \forall i \notin \mathcal{I},
\end{aligned}
\end{equation}
where $\rho \in (0,1]$ denotes the \emph{token manipulation ratio} (TMR), controlling the fraction of vision tokens that can be manipulated.



\subsection{Na\"ive Vision Token Manipulation Attacks}
\label{subsec:attacks}
In this section, we introduce four na\"ive VTM-Attacks that apply different manipulation strategies to $\mathcal{I}$, where $\mathcal{I}$ is selected uniformly at random under the manipulation budget.

\subsubsection{Structure-based Attack}
Structure-based attack alters the positional arrangement of vision tokens while preserving their original values.

\noindent\textbf{Permutation.}
The permutation-based attack applies a derangement $\sigma(\cdot)$ over the selected indices, such that $\hat{{v}}_{\sigma(i)} = {v}_i$, for all $i \in \mathcal{I}$ with $\sigma(i) \neq i$, while preserving all token values. By shuffling token positions, the model processes visual features under mismatched positional context, leading to degraded semantic interpretation.

\subsubsection{Value-based Attack}

Value-based attacks modify the token values while preserving the original ordering.

\noindent\textbf{Masking.}
The masking attack zeroes out the selected tokens:
$\hat{{v}}_i = \mathbf{0}$ for $i \in \mathcal{I}$. This potentially removes the informational content carried by the selected tokens, creating ``cutoffs'' in the visual input.

\noindent\textbf{Gaussian Perturbation.}
The Gaussian perturbation attack injects norm-scaled noise: $\hat{v}_i = v_i + \|v_i\|_2 \epsilon_i$, with $\epsilon_i \sim \mathcal{N}(0, I_d)$ for $i \in \mathcal{I}$. Scaling the noise by $\|v_i\|_2$ ensures that the perturbation magnitude is consistent with the token’s original scale, making it less conspicuous. As a result, each selected vision token undergoes a stochastic perturbation in the embedding space, affecting both its magnitude and direction.

\noindent\textbf{Sign Flip.}
The sign flip attack modifies the selected tokens as $\hat{v}_i = -v_i$ for $i \in \mathcal{I}$. This preserves the token norm, i.e., $\|\hat{v}_i\|_2 = \|v_i\|_2$, while reversing its direction in the embedding space, resulting in a displacement of $\|\hat{v}_i - v_i\|_2 = 2\|v_i\|_2$. As a result, it significantly disrupts the semantic orientation of the vision tokens, making it a particularly destructive manipulation.

\begin{table*}[t]
\centering
\caption{Impact of token manipulation attacks on cloud-edge LVLM inference across four benchmarks. The Attack performance is reported as Accuracy ($\downarrow$) in \%. {\scriptsize\color{deltared}Red} and {\scriptsize\color{deltagreen}green} show absolute change from Benign baselines. Best results are shown in \textbf{bold}.}
\label{tab:main-results}
\small
\setlength{\tabcolsep}{3.2pt}
\renewcommand{\arraystretch}{1.15}
\begin{tabular}{@{}cc|c|cccc|c|c@{}}
\toprule[1.2pt]
\textbf{Benchmark} & \textbf{Model} & \textbf{Benign} & \textbf{Permutation} & \textbf{Masking} & \textbf{Gaussian} & \textbf{Sign Flip} & \textbf{Na\"ive Avg.} & \textbf{Optimization} \\
\midrule
& Qwen2.5-VL-3B-Instruct  & 78.90 & 77.60\dn{-1.30} & 73.50\dn{-5.40} & 66.80\dn{-12.10} & 35.70\dn{-43.20} & 63.40\dn{-15.50} & \textbf{30.50\dn{-48.40}} \\
& Qwen2.5-VL-7B-Instruct  & 83.70 & 83.10\dn{-0.60} & 75.00\dn{-8.70} & 75.50\dn{-8.20} & 19.80\dn{-63.90} & 63.35\dn{-20.35} & \textbf{15.00\dn{-68.70}} \\
& Qwen2.5-VL-72B-Instruct & 84.00 & 83.50\dn{-0.50} & 82.90\dn{-1.10} & 16.40\dn{-67.60} &  \textbf{2.00\dn{-82.00}} & 46.20\dn{-37.80} & \textbf{2.00\dn{-82.00}} \\
\multirow{-2}{*}{\rotatebox[origin=c]{0}{\textbf{OCRBench}}}
& InternVL3.5-4B  & 81.30 & 79.90\dn{-1.40} & 79.70\dn{-1.60} & 79.00\dn{-2.30} & 18.00\dn{-63.30} & 64.15\dn{-17.15} & \textbf{15.70\dn{-65.60}} \\
& InternVL3.5-8B  & 82.70 & 80.90\dn{-1.80} & 81.70\dn{-1.00} & 79.30\dn{-3.40} & 68.60\dn{-14.10} & 77.63\dn{-5.07} & \textbf{65.50\dn{-17.20}} \\
& InternVL3.5-38B & 86.60 & 85.80\dn{-0.80} & 85.70\dn{-0.90} & 84.20\dn{-2.40} & 82.30\dn{-4.30} & 84.50\dn{-2.10} & \textbf{81.20\dn{-5.40}} \\
\midrule
& Qwen2.5-VL-3B-Instruct  & 76.54 & 76.39\dn{-0.15} & 74.14\dn{-2.40} & 49.77\dn{-26.77} &  2.01\dn{-74.53} & 50.58\dn{-25.96} &  \textbf{0.00\dn{-76.54}} \\
& Qwen2.5-VL-7B-Instruct  & 79.87 & 79.56\dn{-0.31} & 24.69\dn{-55.18} & 62.46\dn{-17.41} &  \textbf{0.00\dn{-79.87}} & 41.68\dn{-38.19} &  \textbf{0.00\dn{-79.87}} \\
& Qwen2.5-VL-72B-Instruct & 88.39 & 87.77\dn{-0.62} & 88.31\dn{-0.08} & 32.43\dn{-55.96} &  0.23\dn{-88.16} & 52.19\dn{-36.20} & \textbf{0.08\dn{-88.31}} \\
\multirow{-2}{*}{\rotatebox[origin=c]{0}{\textbf{MMBench}}}
& InternVL3.5-4B  & 81.03 & 79.56\dn{-1.47} & 80.03\dn{-1.00} & 78.25\dn{-2.78} & 27.24\dn{-53.79} & 66.27\dn{-14.76} & \textbf{25.00\dn{-56.03}} \\
& InternVL3.5-8B  & 83.20 & 81.03\dn{-2.17} & 81.42\dn{-1.78} & 80.11\dn{-3.09} & \textbf{79.72\dn{-3.48}} & 80.57\dn{-2.63} & 80.01\dn{-3.19} \\
& InternVL3.5-38B & 86.99 & \textbf{85.75\dn{-1.24}} & 86.53\dn{-0.46} & 86.15\dn{-0.84} & 86.46\dn{-0.53} & 86.22\dn{-0.77} & 86.22\dn{-0.77} \\
\midrule
& Qwen2.5-VL-3B-Instruct  & 60.04 & 59.31\dn{-0.73} & 57.41\dn{-2.63} & 45.53\dn{-14.51} &  \textbf{0.00\dn{-60.04}} & 40.56\dn{-19.48} &  \textbf{0.00\dn{-60.04}} \\
& Qwen2.5-VL-7B-Instruct  & 65.19 & 64.25\dn{-0.94} & 29.34\dn{-35.85} & 59.41\dn{-5.78} &  \textbf{0.00\dn{-65.19}} & 38.25\dn{-26.94} &  \textbf{0.00\dn{-65.19}} \\
& Qwen2.5-VL-72B-Instruct & 69.61 & 68.98\dn{-0.63} & 68.56\dn{-1.05} & 22.61\dn{-47.00} &  0.11\dn{-69.50} & 40.07\dn{-29.54} & \textbf{0.08\dn{-69.53}} \\
\multirow{-2}{*}{\rotatebox[origin=c]{0}{\textbf{HallusionBench}}}
& InternVL3.5-4B  & 69.08 & 68.24\dn{-0.84} & 68.66\dn{-0.42} & 67.72\dn{-1.36} & 36.07\dn{-33.01} & 60.17\dn{-8.91} & \textbf{30.28\dn{-38.80}} \\
& InternVL3.5-8B  & 71.39 & 69.40\dn{-1.99} & 71.82\up{0.43} & \textbf{68.66\dn{-2.73}} & 68.98\dn{-2.41} & 69.72\dn{-1.67} & 69.60\dn{-1.79} \\
& InternVL3.5-38B & 75.07 & 73.19\dn{-1.88} & 74.66\dn{-0.41} & 73.40\dn{-1.67} & 73.40\dn{-1.67} & 73.66\dn{-1.41} & \textbf{52.31\dn{-22.76}} \\
\midrule
& Qwen2.5-VL-3B-Instruct  & 56.40 & 57.10\up{0.70} & 54.10\dn{-2.30} & 43.80\dn{-12.60} & 19.10\dn{-37.30} & 43.53\dn{-12.87} & \textbf{19.00\dn{-37.40}} \\
& Qwen2.5-VL-7B-Instruct  & 45.40 & 44.20\dn{-1.20} & 33.80\dn{-11.60} & 42.50\dn{-2.90} & \textbf{16.80\dn{-28.60}} & 34.33\dn{-11.07} & 17.40\dn{-28.00} \\
& Qwen2.5-VL-72B-Instruct & 47.30 & 27.90\dn{-19.40} &36.40\dn{-10.90} &31.80\dn{-15.50} & 16.70\dn{-30.60} & 28.20\dn{-19.10} & \textbf{15.50\dn{-31.80}} \\
\multirow{-2}{*}{\rotatebox[origin=c]{0}{\textbf{MathVista}}}
& InternVL3.5-4B  & 49.90 & 48.10\dn{-1.80} & 51.80\up{1.90} & 50.60\up{0.70} & 38.00\dn{-11.90} & 47.13\dn{-2.77} & \textbf{35.00\dn{-14.90}} \\
& InternVL3.5-8B  & 48.70 & 47.80\dn{-0.90} & 49.40\up{0.70} & 42.90\dn{-5.80} & 43.70\dn{-5.00} & 45.95\dn{-2.75} & \textbf{42.80\dn{-5.90}} \\
& InternVL3.5-38B & 56.60 & 52.50\dn{-4.10} & 56.30\dn{-0.30} & 54.90\dn{-1.70} & \textbf{50.50\dn{-6.10}} & 53.55\dn{-3.05} & 51.70\dn{-4.90} \\
\bottomrule[1.2pt]
\end{tabular}
\end{table*}

\subsection{Optimization-based Token Selection}
\label{sec:optimized}

The above attacks select tokens uniformly at random, which is suboptimal under a limited manipulation budget. To improve attack effectiveness, we develop an optimization-based token selection strategy that prioritizes the manipulation of the most influential vision tokens.

The objective of the adversary is to alter the model output in Eq.~\ref{eq:overall_obj} by manipulating a subset of vision tokens. However, directly measuring the impact of each token on the output distribution $\mathcal{P}(y_t \mid V, P, Y_{<t}; \boldsymbol{\theta})$ is intractable in the black-box setting. Here, we seek a proxy that captures how perturbations on $V$ affect downstream model behavior.

Self-attention is the primary mechanism for token interaction in Transformer-based LVLMs~\cite{Qwen2.5-VL,wang2025internvl3}, through which vision tokens influence downstream representations. Therefore, maximizing the discrepancy between the self-attention outputs of the original and manipulated vision tokens is expected to induce significant changes in downstream representations.

To enable optimization, we introduce continuous variables ${w} = \{w_1, \dots, w_{L_v}\} \in [0,1]^{L_v}$, where $w_i$ represents the selection weight of vision token $v_i$. The manipulated vision token $\tilde{V}$ is defined as a parameterized transformation of $V$, where ${w}$ controls the extent to which each token is perturbed.
Accordingly, we define the self-attention disruption loss as
\begin{equation}
\begin{aligned}
\mathcal{L}_{\mathrm{SA}}({w}) 
&= \left\| \mathrm{Attn}(\tilde{V}) - \mathrm{Attn}(V) \right\|_F^2, \\
\mathrm{Attn}(V) 
&= \mathrm{softmax}\!\left(\frac{VV^\top}{\sqrt{d}}\right) V,
\end{aligned}
\end{equation}
which quantifies the disruption in self-attention induced by token manipulation.

Maximizing $\mathcal{L}_{\mathrm{SA}}$ encourages selecting tokens whose perturbation most significantly alters the self-attention structure. However, this objective inherently biases the selection toward high-norm tokens, since larger-norm tokens induce greater absolute changes in $VV^\top$ and hence dominate the attention shift (detailed discussion in Appendix~\ref{appendixsec:norm-analysis}~\cite{zhang2026vtm_appendix}).

In practice, such magnitude-driven perturbations can be weakened by downstream normalization layers (e.g., LayerNorm). As a result, larger perturbations do not necessarily translate into stronger effects on the model output.
To mitigate this bias and promote more diverse token selection, we introduce a norm-aware regularization term 
\begin{equation}
\mathcal{L}_{\mathrm{norm}}({w}) 
= \sum_{i=1}^{L_v} w_i \hat{z}_i, 
\quad 
\hat{z}_i = \frac{\|v_i\|_2^2}{\sum_{j=1}^{L_v} \|v_j\|_2^2}.
\end{equation}
This term penalizes selecting high-norm tokens and encourages distributing perturbations across a broader set of tokens. The final objective is given by
\begin{equation}
\max_{{w} \in [0,1]^{L_v}} 
\;\; \mathcal{L}({w}) 
= \mathcal{L}_{\mathrm{SA}}({w}) 
- \lambda \mathcal{L}_{\mathrm{norm}}({w}),
\label{eq:combined-obj}
\end{equation}
where $\lambda$ is a weight that balances the two terms.


To solve the problem in Eq.~\eqref{eq:combined-obj}, we perform projected gradient ascent for $T$ iterations with learning rate $\eta$. At each step, we first update the selection variables via
\begin{equation}
{w} \leftarrow {w} + \eta \nabla_{{w}} \mathcal{L}({w}),
\end{equation}
followed by a projection onto the constraint $w_i \in [0,1]$ via element-wise clipping, which satisfies the budget constraint $\sum_{i=1}^{L_v} w_i = k$ by rescaling ${w}$:
\begin{equation}
{w} \leftarrow {w} \cdot \frac{k}{\sum_{i=1}^{L_v} w_i}.
\end{equation}
The clipping is applied again after rescaling to ensure feasibility. 

After optimization, the discrete token index set is obtained via
\begin{equation}
\mathcal{I} = \mathrm{Top\text{-}k}({w}),
\end{equation}
and the final attack applies the na\"ive VTM-Attack to the selected tokens. We provide the detailed algorithm in Appendix~\ref{appendixsec:algo}~\cite{zhang2026vtm_appendix}.

\begin{figure*}
    \centering
    \includegraphics[width=1\linewidth]{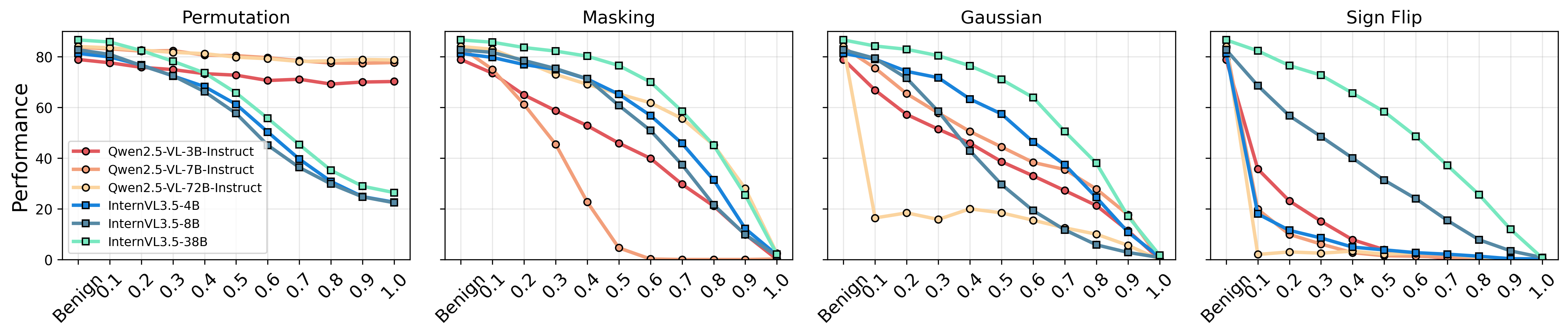}
    \caption{Performance of na\"ive VTM-Attacks under varying TMR on OCRBench.}
    \label{fig:tmr_ocrbench}
\end{figure*}

\section{Experiments}
\label{sec:experiments}
In this section, we describe the experimental setup and evaluate the effectiveness of VTM-Attacks on cloud-edge LVLM inference system.

\subsection{Experimental Setup}

We evaluate six state-of-the-art LVLMs from two representative LVLM families: Qwen2.5-VL-Instruct (including 3B, 7B, and 72B variants)~\cite{Qwen2.5-VL} and InternVL3.5 (including 4B, 8B, and 38B variants)~\cite{wang2025internvl3}. Experiments are conducted on four established vision-language benchmarks: OCRBench~\cite{liu2024ocrbench} assesses scene text understanding, MMBench~\cite{liu2024mmbench} focuses on vision-language instruction following, HallusionBench~\cite{liu2023hallusionbench} targets VQA with an emphasis on visual hallucination and language bias, and MathVista~\cite{lu2023mathvista} evaluates visual mathematical reasoning. Performance is reported as accuracy (\%) for all benchmarks, where lower values under attack indicate stronger attack effectiveness.

We report the \textit{Benign} baseline performance without any attack. For the attack setting, we evaluate multiple VTM-Attack variants, including four na\"ive attack methods, namely \textit{Permutation}, \textit{Masking}, \textit{Gaussian}, and \textit{Sign Flip}, each with randomly selected tokens. We further evaluate an optimization-based token selection strategy introduced in Section~\ref{sec:optimized}, where the selected tokens are manipulated using \textit{Sign Flip}.

Unless otherwise specified, all VTM-Attacks use a token manipulation ratio of $\rho = 0.1$, meaning that only 10\% of vision tokens are modified. For optimization-based token selection, we adopt projected gradient ascent with $T=30$ iterations and a learning rate of $\eta=0.05$. The $\lambda$ is determined using the method described in Appendix~\ref{appendixsec:optimization}~\cite{zhang2026vtm_appendix}.

All benchmark evaluations are conducted using VLMEvalKit~\cite{duan2024vlmevalkit}. Experiments are performed on a large-scale high-performance computing cluster with 20 nodes in parallel, each equipped with four AMD MI250X GPUs (128 GB).

\subsection{Main Results}

In Table~\ref{tab:main-results}, we report the main experimental results. Overall, the benign performance across all benchmarks is consistently high, yet manipulating vision tokens leads to severe degradation in LVLM performance. Na\"ive VTM-Attacks cause substantial drops across models and benchmarks; for example, Qwen2.5-VL-72B-Instruct exhibits average reductions of 37.80\%, 36.20\%, 29.54\%, and 19.10\% on OCRBench, MMBench, HallusionBench, and MathVista, respectively, indicating that simple token-level perturbations are already highly destructive. In particular, on MMBench and HallusionBench, Sign Flip reduces all three Qwen2.5-VL variants to nearly zero accuracy. This suggests that vision tokens serve as a critical semantic interface for vision-language interaction, where small perturbations can disrupt object understanding, spatial relationships, and attribute referencing, ultimately leading to incorrect responses.

A clear difference in robustness is observed across model families. Qwen2.5-VL models exhibit catastrophic degradation under several attacks, whereas InternVL3.5 models are much robust. For instance, on MMBench, Sign Flip reduces InternVL3.5-8B from 83.20\% to 79.72\%, compared to a complete collapse in Qwen2.5-VL-7B-Instruct. This gap persists across benchmarks and attack types, suggesting that robustness is strongly influenced by model architecture and multimodal integration design. Within the InternVL3.5 family, larger models tend to be more robust, while such scaling behavior does not consistently hold for Qwen2.5-VL-Instruct models.

Among the considered attack strategies, Sign Flip consistently produces the strongest degradation. For example, on OCRBench, it reduces Qwen2.5-VL-72B-Instruct from 84.00\% to 2.00\%, while Permutation and Masking result in only minor performance drops. A similar pattern is observed across other benchmarks, where Permutation is generally the least effective due to preserving token values, and Masking produces moderate degradation by removing token content. Gaussian perturbation shows variable behavior and performs effective on some models and benchmarks. Intuitively, Sign Flip preserves token norm while reversing direction, causing large displacement in the representation space and severely disrupting token interactions.

Finally, the optimization-based token selection strategy further amplifies attack effectiveness over random selection. For example, on OCRBench, optimization-based method reduces random selection-based Sign Flip for Qwen2.5-VL-7B-Instruct model from 19.80\% to 15.00\%, and on HallusionBench, InternVL3.5-4B and InternVL3.5-38B, optimization-based method improves +5.79\% and +21.09\%, respectively. These results demonstrate that the proposed selection method can effectively identify more vulnerable vision tokens.

\subsection{Ablation Studies}

\noindent\textbf{Influence on Token Manipulation Ratio $\rho$.}
Fig.~\ref{fig:tmr_ocrbench} illustrates the impact of varying the TMR $\rho$ on OCRBench with different attack strategies. Overall, increasing $\rho$ leads to progressively stronger performance degradation across all attacks. Specifically, different attack types exhibit distinct sensitivity to $\rho$. Permutation introduces relatively mild degradation, indicating that preserving token values limits its impact even when more tokens are affected. In contrast, value-based attacks, including Masking, Gaussian perturbation, and Sign Flip, show significantly stronger dependence on $\rho$, with performance dropping rapidly as $\rho$ increases. Among them, Sign Flip is consistently the most aggressive, often causing near-complete failure at a large $\rho$.
We also observe a clear difference across model families. Qwen models degrade sharply even at low $\rho$, while InternVL models exhibit more gradual degradation as $\rho$ increases, suggesting stronger robustness to partial token corruption. These results highlight that both the manipulation budget and the attack type jointly determine the effectiveness of VTM-Attacks in cloud-edge LVLM inference. We provide additional results of varying TMR on other benchmarks in Appendix~\ref{appendixsec:add_abl}~\cite{zhang2026vtm_appendix}.

\begin{figure}
    \centering
    \includegraphics[width=0.80\linewidth]{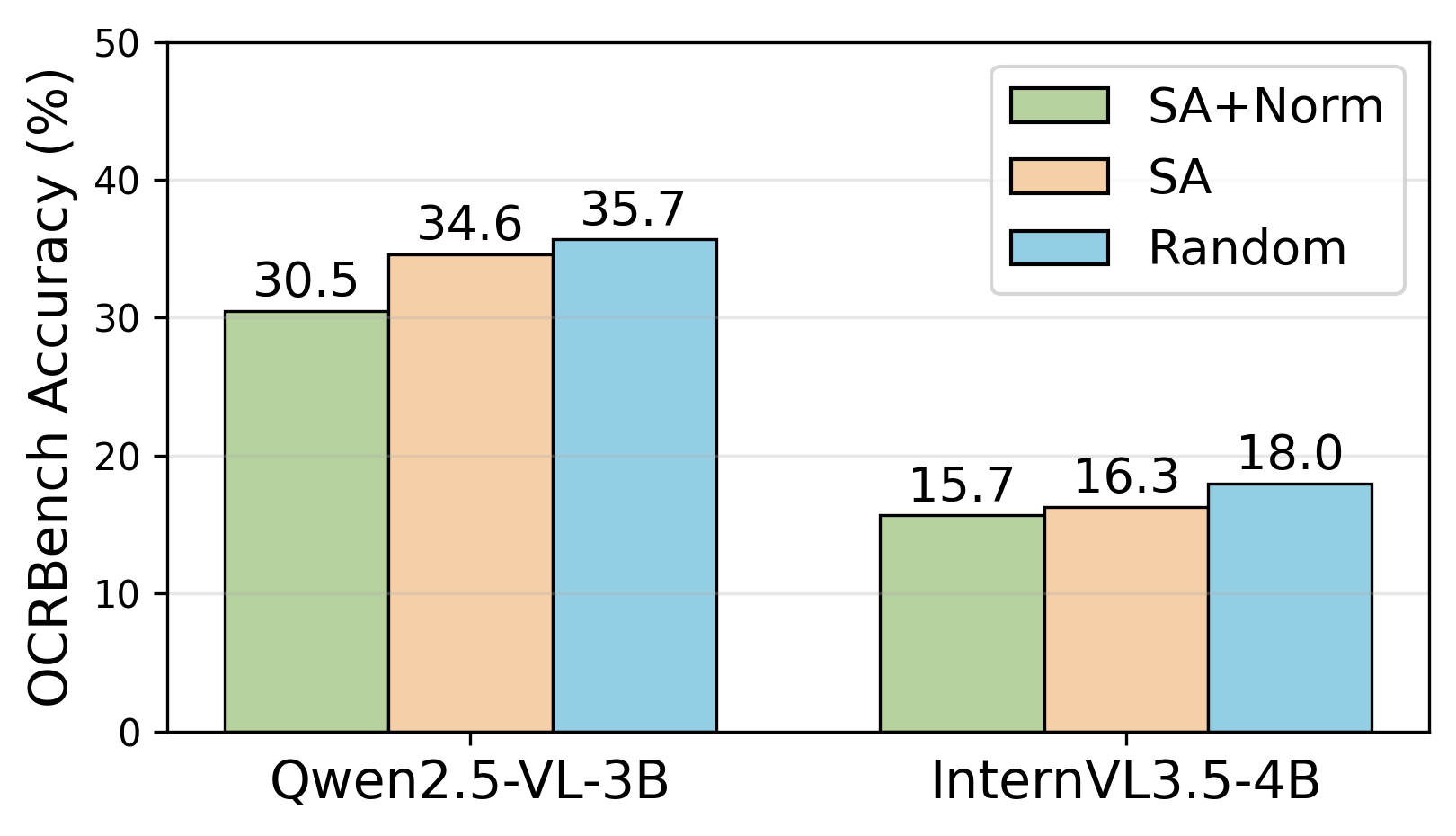}
    \caption{Effectiveness of optimization objectives on OCRBench.}
    \label{fig:loss_ablation_ocr}
\end{figure}

\noindent\textbf{Effectiveness of Optimization Objectives.}
To further evaluate the contribution of different optimization objectives. We conduct ablation studies on OCRBench by comparing three token selection strategies: random selection, self-attention disruption only (SA), and the full objective with additional norm-aware regularization (SA+Norm). As shown in Fig.~\ref{fig:loss_ablation_ocr}, the full objective consistently achieves the strongest attack effectiveness, yielding the lowest accuracy across both models. For example, on Qwen2.5-VL-3B, the accuracy decreases from 35.70\% under random selection to 34.60\% with SA, and further to 30.50\% with SA+Norm. A similar trend is observed on InternVL3.5-4B, where performance drops from 18.00\% to 16.30\% and then to 15.70\%. These results demonstrate that self-attention disruption is an effective criterion for identifying influential tokens, while incorporating norm-aware regularization further improves the selection by reducing the bias toward high-norm tokens, leading to more effective attacks.

\begin{figure}[t]
    \centering
    \includegraphics[width=0.85\linewidth]{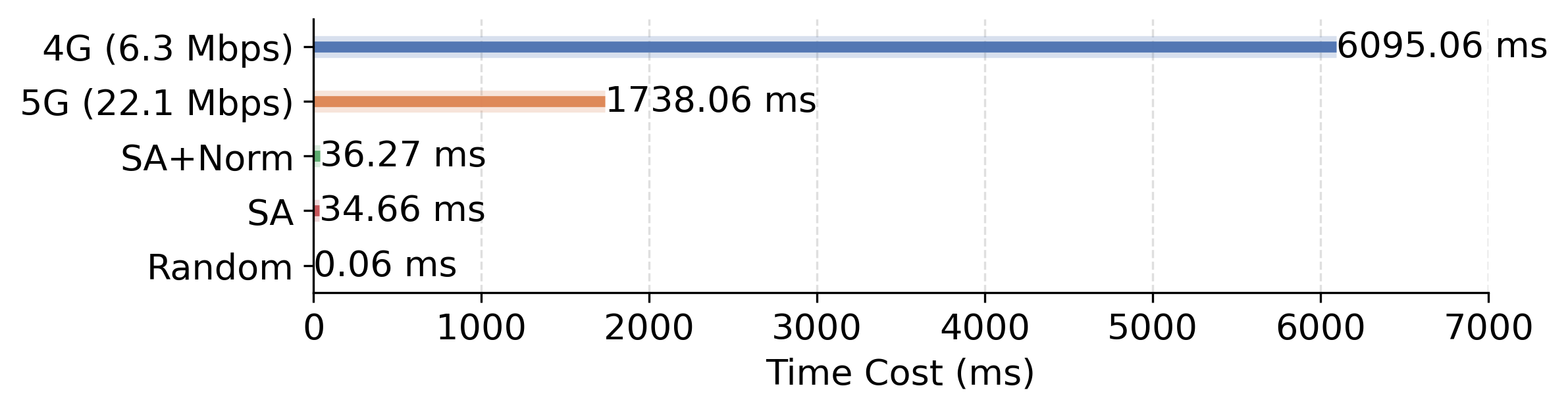}
    \caption{Comparison of communication latency and VTM-Attacks time cost.}
    \label{fig:latency}
\end{figure}

\subsection{Efficiency of TMA-Attacks}
In Fig.~\ref{fig:latency}, we compare the time cost of token manipulation with the communication latency under typical network settings~\cite{jiang2025hyperion}. For Qwen2.5-VL-3B with 1,225 visual tokens (4.8 MB in FP16, averaged over OCRBench), transmission over 4G and 5G takes 6,095.06 ms and 1,738.06 ms, respectively, whereas all TMA-Attack variants operate at the millisecond scale. Specifically, the optimization-based strategy (SA+Norm) takes only 36.27 ms, the SA-only variant takes 34.66 ms, and the random selection strategy takes as little as 0.06 ms. Overall, the attack introduces negligible additional latency to the cloud-edge inference pipeline.

\section{Conclusion}\label{sec:conclusion}
In this work, we identify the vision token transmission interface in cloud-edge LVLM inference as a critical yet overlooked attack surface. Based on this insight, we propose a family of VTM-Attacks, comprising four na\"ive attack strategies and an optimization-based token selection method driven by self-attention disruption and norm-aware regularization. Experiments on six LVLMs across four benchmarks show that manipulating a small portion of vision tokens can significantly reduce accuracy, while introducing negligible attack latency compared to transmission overhead. These findings underscore the urgent need for robust defenses at the communication interface of cloud-edge LVLM inference systems.

\section*{Acknowledgment}
This material is based upon work co-supported by the U.S. Department of Energy, Office of Science, Office of Advanced Scientific Computing Research under Contract No. DE-AC05-00OR22725. This manuscript has been co-authored by UT-Battelle, LLC under Contract No. DE-AC05-00OR22725 with the U.S. Department of Energy. The United States Government retains and the publisher, by accepting the article for publication, acknowledges that the United States Government retains a non-exclusive, paid-up, irrevocable, world-wide license to publish or reproduce the published form of this manuscript, or allow others to do so, for United States Government purposes. The Department of Energy will provide public access to these results of federally sponsored research in accordance with the DOE Public Access Plan (http://energy.gov/downloads/doe-public-access-plan).

{
\small
\bibliographystyle{plain}
\bibliography{main}
}

\appendix

This appendix provides additional experimental results, theoretical analysis, and algorithmic details. The content is organized as follows:

\begin{itemize}
    \item \textbf{Appendix Section~\ref{appendixsec:add_abl}}: Effects of Token Manipulation Ratio on Na\"ive Vision Token Manipulation Attacks

    \item \textbf{Appendix Section~\ref{appendixsec:algo}}: Optimization-based Token Selection Algorithm

    \item \textbf{Appendix Section~\ref{appendixsec:norm-analysis}}: Norm Bias in Self-Attention Disruption

    \item \textbf{Appendix Section~\ref{appendixsec:optimization}}: Adaptive $\lambda$ Calibration


\end{itemize}

\subsection{Effects of Token Manipulation Ratio on Na\"ive Vision Token Manipulation Attacks}\label{appendixsec:add_abl}

\begin{table*}[h]
\centering
\caption{Performance under different token manipulation attacks with varying TMR on OCRBench.}
\label{tab:tmr_ocrbench}
\small
\setlength{\tabcolsep}{3pt}
\renewcommand{\arraystretch}{1.1}

\begin{tabular}{l|c|cccccccccc}
\toprule
Model & Benign & 0.1 & 0.2 & 0.3 & 0.4 & 0.5 & 0.6 & 0.7 & 0.8 & 0.9 & 1.0 \\
\midrule

\multicolumn{12}{c}{\textbf{Permutation}} \\
\midrule
Qwen2.5-VL-3B-Instruct  & 78.90 & 77.60 & 75.80 & 74.90 & 73.30 & 72.70 & 70.60 & 71.10 & 69.20 & 70.00 & 70.20 \\
Qwen2.5-VL-7B-Instruct  & 83.70 & 83.10 & 82.20 & 82.40 & 80.60 & 80.40 & 79.60 & 78.40 & 77.50 & 77.40 & 77.70 \\
Qwen2.5-VL-72B-Instruct & 84.00 & 83.50 & 82.60 & 81.70 & 81.20 & 79.80 & 79.20 & 78.10 & 78.40 & 78.90 & 78.70 \\
InternVL3.5-4B          & 81.30 & 79.90 & 76.70 & 72.40 & 68.20 & 61.20 & 50.30 & 39.60 & 30.80 & 24.80 & 22.60 \\
InternVL3.5-8B          & 82.70 & 80.90 & 76.50 & 72.50 & 66.20 & 57.60 & 45.10 & 36.30 & 29.80 & 24.70 & 22.40 \\
InternVL3.5-38B         & 86.60 & 85.80 & 82.40 & 78.20 & 73.60 & 65.70 & 55.70 & 45.30 & 35.20 & 28.90 & 26.40 \\

\midrule
\multicolumn{12}{c}{\textbf{Masking}} \\
\midrule
Qwen2.5-VL-3B-Instruct  & 78.90 & 73.50 & 64.90 & 58.70 & 52.80 & 45.90 & 39.90 & 29.70 & 21.10 & 9.90 & 0.10 \\
Qwen2.5-VL-7B-Instruct  & 83.70 & 75.00 & 61.20 & 45.50 & 22.70 & 4.60 & 0.20 & 0.00 & 0.00 & 0.00 & 0.20 \\
Qwen2.5-VL-72B-Instruct & 84.00 & 82.90 & 78.10 & 73.00 & 69.10 & 65.30 & 61.80 & 55.60 & 45.20 & 28.00 & 2.40 \\
InternVL3.5-4B          & 81.30 & 79.70 & 76.90 & 75.00 & 71.20 & 65.20 & 56.80 & 45.80 & 31.40 & 12.20 & 1.80 \\
InternVL3.5-8B          & 82.70 & 81.70 & 78.50 & 75.30 & 71.30 & 60.80 & 50.90 & 37.40 & 21.60 & 9.80 & 1.40 \\
InternVL3.5-38B         & 86.60 & 85.70 & 83.60 & 82.20 & 80.20 & 76.50 & 70.00 & 58.50 & 45.10 & 25.40 & 2.10 \\

\midrule
\multicolumn{12}{c}{\textbf{Gaussian Perturbation}} \\
\midrule
Qwen2.5-VL-3B-Instruct  & 78.90 & 66.80 & 57.20 & 51.40 & 45.90 & 38.60 & 33.00 & 27.20 & 21.30 & 11.40 & 0.10 \\
Qwen2.5-VL-7B-Instruct  & 83.70 & 75.50 & 65.40 & 57.80 & 50.50 & 44.40 & 38.30 & 35.50 & 27.80 & 17.70 & 0.10 \\
Qwen2.5-VL-72B-Instruct & 84.00 & 16.40 & 18.40 & 15.80 & 20.00 & 18.40 & 15.50 & 12.50 & 10.00 & 5.60 & 0.00 \\
InternVL3.5-4B          & 81.30 & 79.00 & 74.20 & 71.70 & 63.30 & 57.40 & 46.40 & 37.40 & 24.50 & 10.80 & 1.00 \\
InternVL3.5-8B          & 82.70 & 79.30 & 71.50 & 58.40 & 42.90 & 29.60 & 19.30 & 11.70 & 5.80 & 2.80 & 0.80 \\
InternVL3.5-38B         & 86.60 & 84.20 & 82.90 & 80.40 & 76.40 & 71.00 & 63.90 & 50.50 & 38.00 & 17.10 & 1.70 \\

\midrule
\multicolumn{12}{c}{\textbf{Sign Flip}} \\
\midrule
Qwen2.5-VL-3B-Instruct  & 78.90 & 35.70 & 23.10 & 15.10 & 7.80 & 3.90 & 1.70 & 0.40 & 0.20 & 0.00 & 0.10 \\
Qwen2.5-VL-7B-Instruct  & 83.70 & 19.80 & 9.90 & 6.10 & 2.70 & 1.50 & 1.30 & 0.90 & 0.30 & 0.00 & 0.00 \\
Qwen2.5-VL-72B-Instruct & 84.00 & 2.00 & 3.00 & 2.50 & 3.30 & 2.10 & 2.50 & 2.10 & 0.90 & 0.40 & 0.00 \\
InternVL3.5-4B          & 81.30 & 18.00 & 11.50 & 8.50 & 4.90 & 3.80 & 2.70 & 2.10 & 1.30 & 0.30 & 0.30 \\
InternVL3.5-8B          & 82.70 & 68.60 & 56.70 & 48.50 & 40.00 & 31.30 & 24.00 & 15.40 & 7.80 & 3.40 & 0.80 \\
InternVL3.5-38B         & 86.60 & 82.30 & 76.50 & 72.70 & 65.60 & 58.30 & 48.60 & 37.10 & 25.60 & 11.90 & 0.60 \\

\bottomrule
\end{tabular}
\end{table*}

In this section, we analyze how attack effectiveness changes as the TMR increases. We vary $\rho \in \{0.1,0.2,\dots,1.0\}$ and report the corresponding accuracy for each attack type and model in Tables~\ref{tab:tmr_ocrbench}--\ref{tab:tmr_mathvista}. Overall, a larger TMR generally leads to more severe degradation, but the rate of degradation differs substantially across attack primitives and model families. s

\begin{table*}[h]
\centering
\caption{Performance under different token manipulation attacks with varying TMR on MMBench.}
\label{tab:tmr_mmbench}
\small
\setlength{\tabcolsep}{3pt}
\renewcommand{\arraystretch}{1.1}

\begin{tabular}{l|c|cccccccccc}
\toprule
Model & Benign & 0.1 & 0.2 & 0.3 & 0.4 & 0.5 & 0.6 & 0.7 & 0.8 & 0.9 & 1.0 \\
\midrule

\multicolumn{12}{c}{\textbf{Permutation}} \\
\midrule
Qwen2.5-VL-3B-Instruct  & 76.54 & 76.39 & 76.39 & 76.00 & 75.07 & 75.23 & 74.53 & 74.61 & 75.15 & 73.91 & 73.99 \\
Qwen2.5-VL-7B-Instruct  & 79.87 & 79.56 & 79.25 & 79.02 & 78.86 & 78.63 & 78.56 & 78.09 & 78.09 & 78.01 & 77.39 \\
Qwen2.5-VL-72B-Instruct & 88.39 & 87.77 & 88.23 & 87.30 & 87.07 & 87.22 & 86.91 & 86.84 & 86.14 & 86.60 & 86.14 \\
InternVL3.5-4B          & 81.03 & 79.56 & 78.71 & 77.63 & 76.62 & 74.76 & 73.52 & 72.21 & 71.28 & 70.66 & 70.12 \\
InternVL3.5-8B          & 83.20 & 81.03 & 79.56 & 78.17 & 76.39 & 75.77 & 72.98 & 72.91 & 71.28 & 69.89 & 69.73 \\
InternVL3.5-38B         & 86.99 & 85.75 & 84.21 & 82.19 & 81.42 & 80.41 & 79.64 & 78.17 & 76.85 & 77.01 & 75.77 \\

\midrule
\multicolumn{12}{c}{\textbf{Masking}} \\
\midrule
Qwen2.5-VL-3B-Instruct  & 76.54 & 74.14 & 71.82 & 70.89 & 67.49 & 61.14 & 40.55 & 2.47 & 0.00 & 0.00 & 0.00 \\
Qwen2.5-VL-7B-Instruct  & 79.87 & 24.69 & 7.35 & 2.47 & 0.61 & 0.07 & 0.07 & 0.00 & 0.00 & 0.00 & 0.00 \\
Qwen2.5-VL-72B-Instruct & 88.39 & 88.31 & 87.92 & 87.69 & 86.37 & 81.81 & 75.15 & 76.70 & 80.65 & 70.89 & 13.31 \\
InternVL3.5-4B          & 81.03 & 80.03 & 79.41 & 78.79 & 77.78 & 76.00 & 74.53 & 73.06 & 70.20 & 60.91 & 17.87 \\
InternVL3.5-8B          & 83.20 & 81.42 & 81.26 & 79.64 & 78.79 & 77.55 & 77.01 & 73.99 & 69.42 & 61.22 & 16.40 \\
InternVL3.5-38B         & 86.99 & 86.53 & 85.99 & 85.91 & 84.98 & 84.75 & 83.43 & 82.04 & 79.48 & 73.45 & 19.42 \\

\midrule
\multicolumn{12}{c}{\textbf{Gaussian Perturbation}} \\
\midrule
Qwen2.5-VL-3B-Instruct  & 76.54 & 49.77 & 43.89 & 38.78 & 31.89 & 26.63 & 17.26 & 11.15 & 5.34 & 2.17 & 1.01 \\
Qwen2.5-VL-7B-Instruct  & 79.87 & 62.46 & 64.55 & 54.95 & 48.68 & 37.69 & 25.69 & 16.79 & 1.70 & 0.85 & 0.46 \\
Qwen2.5-VL-72B-Instruct & 88.39 & 32.43 & 5.88 & 0.85 & 0.30 & 0.30 & 0.30 & 0.07 & 0.23 & 0.00 & 0.00 \\
InternVL3.5-4B          & 81.03 & 78.25 & 77.48 & 76.24 & 74.07 & 73.30 & 69.89 & 68.19 & 63.70 & 54.18 & 13.31 \\
InternVL3.5-8B          & 83.20 & 80.11 & 77.94 & 75.08 & 68.42 & 62.54 & 51.63 & 48.60 & 41.02 & 28.79 & 6.03 \\
InternVL3.5-38B         & 86.99 & 86.15 & 85.45 & 85.60 & 84.52 & 84.37 & 82.66 & 80.11 & 76.24 & 68.96 & 16.18 \\

\midrule
\multicolumn{12}{c}{\textbf{Sign Flip}} \\
\midrule
Qwen2.5-VL-3B-Instruct  & 76.54 & 2.01 & 1.23 & 1.00 & 0.77 & 0.38 & 0.46 & 0.30 & 0.23 & 0.15 & 0.07 \\
Qwen2.5-VL-7B-Instruct  & 79.87 & 0.00 & 0.00 & 0.00 & 0.00 & 0.00 & 0.00 & 0.00 & 0.00 & 0.00 & 0.00 \\
Qwen2.5-VL-72B-Instruct & 88.39 & 0.23 & 0.23 & 0.38 & 0.23 & 0.23 & 0.30 & 0.23 & 0.23 & 0.23 & 0.30 \\
InternVL3.5-4B          & 81.03 & 27.24 & 22.67 & 19.65 & 18.26 & 16.95 & 16.25 & 14.24 & 12.46 & 9.52 & 1.70 \\
InternVL3.5-8B          & 83.20 & 79.72 & 76.93 & 73.07 & 68.96 & 63.62 & 56.97 & 49.07 & 44.73 & 29.64 & 8.04 \\
InternVL3.5-38B         & 86.99 & 86.46 & 85.76 & 84.60 & 81.42 & 74.85 & 69.12 & 59.83 & 56.57 & 47.75 & 10.60 \\

\bottomrule
\end{tabular}
\end{table*}

\begin{table*}[h]
\centering
\caption{Performance under different token manipulation attacks with varying TMR on HallusionBench.}
\label{tab:tmr_hallusion}
\small
\setlength{\tabcolsep}{3pt}
\renewcommand{\arraystretch}{1.1}

\begin{tabular}{l|c|cccccccccc}
\toprule
Model & Benign & 0.1 & 0.2 & 0.3 & 0.4 & 0.5 & 0.6 & 0.7 & 0.8 & 0.9 & 1.0 \\
\midrule

\multicolumn{12}{c}{\textbf{Permutation}} \\
\midrule
Qwen2.5-VL-3B-Instruct  & 60.04 & 59.31 & 58.99 & 60.25 & 58.89 & 58.25 & 57.94 & 58.15 & 56.78 & 57.31 & 57.41 \\
Qwen2.5-VL-7B-Instruct  & 65.19 & 64.25 & 63.51 & 63.41 & 62.04 & 62.15 & 61.93 & 61.72 & 62.15 & 61.93 & 61.41 \\
Qwen2.5-VL-72B-Instruct & 69.61 & 68.98 & 68.03 & 68.03 & 67.61 & 66.56 & 66.25 & 65.51 & 65.09 & 64.25 & 64.77 \\
InternVL3.5-4B          & 69.08 & 68.24 & 67.61 & 67.40 & 67.19 & 64.98 & 65.72 & 61.72 & 60.78 & 60.15 & 60.15 \\
InternVL3.5-8B          & 71.39 & 69.40 & 68.45 & 66.25 & 64.98 & 64.98 & 62.25 & 59.52 & 58.25 & 57.31 & 58.15 \\
InternVL3.5-38B         & 75.07 & 73.19 & 72.13 & 72.24 & 68.87 & 68.66 & 66.04 & 62.46 & 61.83 & 60.67 & 60.88 \\

\midrule
\multicolumn{12}{c}{\textbf{Masking}} \\
\midrule
Qwen2.5-VL-3B-Instruct  & 60.04 & 57.41 & 55.10 & 51.74 & 46.69 & 38.38 & 26.29 & 7.47 & 1.26 & 0.74 & 0.11 \\
Qwen2.5-VL-7B-Instruct  & 65.19 & 29.34 & 16.51 & 4.21 & 0.63 & 0.00 & 0.00 & 0.00 & 0.00 & 0.00 & 0.00 \\
Qwen2.5-VL-72B-Instruct & 69.61 & 68.56 & 68.03 & 65.62 & 65.19 & 63.20 & 65.72 & 62.67 & 60.99 & 52.89 & 46.37 \\
InternVL3.5-4B          & 69.08 & 68.66 & 68.77 & 67.09 & 67.09 & 64.88 & 64.98 & 63.72 & 62.46 & 56.15 & 54.99 \\
InternVL3.5-8B          & 71.39 & 71.82 & 69.93 & 70.45 & 68.35 & 65.72 & 63.62 & 62.25 & 59.20 & 54.99 & 53.10 \\
InternVL3.5-38B         & 75.07 & 74.66 & 74.13 & 73.08 & 72.56 & 70.45 & 70.66 & 67.93 & 65.62 & 60.67 & 54.26 \\

\midrule
\multicolumn{12}{c}{\textbf{Gaussian Perturbation}} \\
\midrule
Qwen2.5-VL-3B-Instruct  & 60.04 & 45.53 & 38.49 & 28.92 & 20.08 & 13.88 & 10.41 & 6.62 & 5.36 & 4.73 & 4.42 \\
Qwen2.5-VL-7B-Instruct  & 65.19 & 59.41 & 51.74 & 38.07 & 17.25 & 10.30 & 6.62 & 3.15 & 2.63 & 2.10 & 1.05 \\
Qwen2.5-VL-72B-Instruct & 69.61 & 22.61 & 6.52 & 0.32 & 0.21 & 0.11 & 0.00 & 0.00 & 0.00 & 0.00 & 0.00 \\
InternVL3.5-4B          & 69.08 & 67.72 & 67.30 & 67.72 & 64.98 & 64.04 & 62.46 & 63.20 & 58.68 & 57.20 & 55.10 \\
InternVL3.5-8B          & 71.39 & 68.66 & 68.35 & 65.83 & 62.88 & 59.73 & 53.84 & 49.63 & 40.69 & 29.34 & 18.19 \\
InternVL3.5-38B         & 75.07 & 73.40 & 73.40 & 71.50 & 71.40 & 68.77 & 69.09 & 64.04 & 59.83 & 55.52 & 51.84 \\

\midrule
\multicolumn{12}{c}{\textbf{Sign Flip}} \\
\midrule
Qwen2.5-VL-3B-Instruct  & 60.04 & 0.00 & 0.21 & 0.00 & 0.00 & 0.21 & 0.32 & 0.11 & 0.00 & 0.00 & 0.00 \\
Qwen2.5-VL-7B-Instruct  & 65.19 & 0.00 & 0.00 & 0.00 & 0.00 & 0.00 & 0.00 & 0.11 & 0.00 & 0.00 & 0.00 \\
Qwen2.5-VL-72B-Instruct & 69.61 & 0.11 & 0.00 & 0.00 & 0.00 & 0.00 & 0.00 & 0.00 & 0.00 & 0.00 & 0.00 \\
InternVL3.5-4B          & 69.08 & 36.07 & 23.66 & 18.09 & 14.72 & 14.20 & 12.30 & 10.52 & 10.62 & 10.83 & 9.36 \\
InternVL3.5-8B          & 71.39 & 68.98 & 67.61 & 66.77 & 63.72 & 60.36 & 59.20 & 53.52 & 51.95 & 45.95 & 33.12 \\
InternVL3.5-38B         & 75.07 & 73.40 & 72.77 & 71.50 & 70.56 & 66.98 & 64.56 & 60.15 & 54.68 & 52.16 & 46.79 \\

\bottomrule
\end{tabular}
\end{table*}

\begin{table*}[h]
\centering
\caption{Performance under different token manipulation attacks with varying TMR on MathVista.}
\label{tab:tmr_mathvista}
\small
\setlength{\tabcolsep}{3pt}
\renewcommand{\arraystretch}{1.1}

\begin{tabular}{l|c|cccccccccc}
\toprule
Model & Benign & 0.1 & 0.2 & 0.3 & 0.4 & 0.5 & 0.6 & 0.7 & 0.8 & 0.9 & 1.0 \\
\midrule

\multicolumn{12}{c}{\textbf{Permutation}} \\
\midrule
Qwen2.5-VL-3B-Instruct  & 56.40 & 57.10 & 56.40 & 56.90 & 56.60 & 55.10 & 54.90 & 55.40 & 54.30 & 52.90 & 52.30 \\
Qwen2.5-VL-7B-Instruct  & 45.40 & 44.20 & 44.20 & 44.80 & 45.30 & 42.90 & 44.30 & 44.40 & 44.40 & 43.90 & 42.00 \\
Qwen2.5-VL-72B-Instruct & 47.30 & 27.90 & 19.30 & 20.05 & 22.80 & 19.60 & 19.50 & 24.10 & 27.60 & 22.80 & 26.20 \\
InternVL3.5-4B          & 49.90 & 48.10 & 45.20 & 45.10 & 41.70 & 40.40 & 36.30 & 34.90 & 34.20 & 32.00 & 32.90 \\
InternVL3.5-8B          & 48.70 & 47.80 & 46.50 & 44.80 & 40.80 & 37.80 & 35.90 & 34.20 & 33.80 & 32.70 & 31.50 \\
InternVL3.5-38B         & 56.60 & 52.50 & 51.10 & 50.40 & 46.60 & 43.10 & 41.40 & 38.90 & 35.90 & 36.00 & 35.80 \\

\midrule
\multicolumn{12}{c}{\textbf{Masking}} \\
\midrule
Qwen2.5-VL-3B-Instruct  & 56.40 & 54.10 & 50.40 & 46.50 & 42.50 & 37.10 & 33.00 & 25.20 & 20.30 & 19.20 & 16.60 \\
Qwen2.5-VL-7B-Instruct  & 45.40 & 33.80 & 31.00 & 24.80 & 19.90 & 16.20 & 16.80 & 17.70 & 18.70 & 18.00 & 17.20 \\
Qwen2.5-VL-72B-Instruct & 47.30 & 36.40 & 37.90 & 35.80 & 42.50 & 38.50 & 28.60 & 35.20 & 31.40 & 23.70 & 15.80 \\
InternVL3.5-4B          & 49.90 & 51.80 & 49.30 & 49.70 & 47.60 & 47.60 & 44.40 & 41.20 & 37.30 & 30.40 & 23.20 \\
InternVL3.5-8B          & 48.70 & 49.40 & 48.70 & 48.70 & 47.90 & 47.60 & 44.20 & 41.40 & 38.50 & 36.00 & 26.30 \\
InternVL3.5-38B         & 56.60 & 56.30 & 56.49 & 54.90 & 55.20 & 52.50 & 48.69 & 45.40 & 42.19 & 35.30 & 28.90 \\

\midrule
\multicolumn{12}{c}{\textbf{Gaussian Perturbation}} \\
\midrule
Qwen2.5-VL-3B-Instruct  & 56.40 & 43.80 & 36.80 & 35.60 & 30.50 & 27.30 & 25.10 & 24.60 & 21.90 & 20.50 & 20.40 \\
Qwen2.5-VL-7B-Instruct  & 45.40 & 42.50 & 37.40 & 32.50 & 26.00 & 21.80 & 21.20 & 23.20 & 20.40 & 21.90 & 23.10 \\
Qwen2.5-VL-72B-Instruct & 47.30 & 31.80 & 25.50 & 21.60 & 21.50 & 20.40 & 22.10 & 18.90 & 20.50 & 19.00 & 19.10 \\
InternVL3.5-4B          & 49.90 & 50.60 & 49.40 & 50.40 & 46.40 & 42.90 & 39.70 & 35.10 & 34.20 & 29.00 & 25.20 \\
InternVL3.5-8B          & 48.70 & 42.90 & 38.30 & 35.90 & 34.00 & 30.60 & 27.70 & 24.40 & 22.90 & 21.00 & 19.30 \\
InternVL3.5-38B         & 56.60 & 54.90 & 55.20 & 52.50 & 54.30 & 50.60 & 47.90 & 45.00 & 41.09 & 36.90 & 30.50 \\

\midrule
\multicolumn{12}{c}{\textbf{Sign Flip}} \\
\midrule
Qwen2.5-VL-3B-Instruct  & 56.40 & 19.10 & 19.30 & 20.80 & 19.40 & 18.00 & 17.80 & 18.40 & 18.50 & 18.30 & 16.30 \\
Qwen2.5-VL-7B-Instruct  & 45.40 & 16.80 & 15.60 & 15.20 & 15.90 & 16.10 & 15.70 & 15.00 & 15.30 & 14.90 & 16.80 \\
Qwen2.5-VL-72B-Instruct & 47.30 & 16.70 & 16.70 & 18.00 & 17.40 & 17.29 & 15.29 & 16.70 & 16.40 & 15.80 & 18.20 \\
InternVL3.5-4B          & 49.90 & 38.00 & 32.30 & 30.70 & 29.70 & 27.50 & 25.40 & 23.40 & 23.10 & 21.30 & 19.70 \\
InternVL3.5-8B          & 48.70 & 43.70 & 41.90 & 40.00 & 39.60 & 37.90 & 34.20 & 31.00 & 27.90 & 23.80 & 20.50 \\
InternVL3.5-38B         & 56.60 & 50.50 & 51.50 & 50.70 & 50.70 & 50.00 & 46.80 & 45.40 & 42.40 & 35.50 & 27.70 \\

\bottomrule
\end{tabular}
\end{table*}

\subsection{Algorithm}\label{appendixsec:algo}

\begin{algorithm}[h]
\caption{Optimization-based Token Selection}
\label{alg:opt_select}
\begin{algorithmic}[1]
\Require Vision tokens $V = [v_1,\dots,v_{L_v}]^\top$;
token manipulation ratio $\rho$;
number of iterations $T$;
learning rate $\eta$;
regularization weight $\lambda$.

\State Set the manipulation budget:
$k \leftarrow \lfloor \rho L_v \rfloor$.

\State Initialize the selection variables:
$w_i^{(0)} \leftarrow \frac{k}{L_v}$, for all $i \in [1,L_v]$.

\For{$t \in [1,T]$}
    \State Construct the manipulated vision tokens $\tilde{V}^{(t)}$ from $V$ and $w^{(t-1)}$.
    
    \State Compute the self-attention disruption loss:
    \[
    \mathcal{L}_{\mathrm{SA}}(w^{(t-1)})
    =
    \left\|
    \mathrm{Attn}(\tilde{V}^{(t)}) - \mathrm{Attn}(V)
    \right\|_F^2
    \]
    
    \State Compute the norm-aware regularization:
    \[
    \mathcal{L}_{\mathrm{norm}}(w^{(t-1)})
    =
    \sum_{i=1}^{L_v} w_i^{(t-1)} \hat{z}_i
    \]
    
    \State Compute the objective:
    \[
    \mathcal{L}(w^{(t-1)})
    =
    \mathcal{L}_{\mathrm{SA}}(w^{(t-1)})
    -
    \lambda \mathcal{L}_{\mathrm{norm}}(w^{(t-1)})
    \]
    
    \State Update the selection variables:
    \[
    w^{(t)}
    \leftarrow
    w^{(t-1)} + \eta \nabla_{w}\mathcal{L}(w^{(t-1)})
    \]
    
    \State Clip $w^{(t)}$ element wise to $[0,1]$.
    
    \State Rescale $w^{(t)}$ to satisfy the budget constraint:
    \[
    w^{(t)}
    \leftarrow
    w^{(t)} \cdot \frac{k}{\sum_{i=1}^{L_v} w_i^{(t)}}
    \]
    
    \State Clip $w^{(t)}$ element wise to $[0,1]$ again.
\EndFor

\State Obtain the discrete token set:
$\mathcal{I} \leftarrow \mathrm{Top\text{-}k}(w^{(T)})$.

\State Apply the chosen token manipulation to the selected tokens in $\mathcal{I}$.

\State \Return $\mathcal{I}$
\end{algorithmic}
\end{algorithm}

The detailed optimization procedure is summarized in Algorithm~\ref{alg:opt_select}. During optimization, we construct the manipulated vision tokens using a differentiable relaxation of the token manipulation operator. For example, for sign flip, we adopt a continuous relaxation such that at iteration $t$, each token is transformed as
\begin{equation}
\tilde v_i^{(t)} = (1 - 2w_i^{(t-1)}) v_i,
\end{equation}
where $w_i^{(t-1)} \in [0,1]$ controls the degree of transformation. This formulation provides a differentiable approximation to discrete sign flipping, where $w_i=0$ keeps the original token unchanged and $w_i=1$ corresponds to full sign inversion. The resulting manipulated token matrix $\tilde V^{(t)}$ is then used to evaluate the optimization objective.

\subsection{Norm Bias in Self-Attention Disruption}
\label{appendixsec:norm-analysis}

We consider perturbations that modify a selected token either by value transformation or by token replacement. Formally, for each selected token $v_i$, the perturbed token is written as
\begin{equation}
\tilde v_i = \phi_i(V),
\end{equation}
where $\phi_i$ denotes a transformation operator. For value-based attacks such as masking, Gaussian perturbation, and sign flipping, this reduces to the additive form $\tilde v_i = v_i + \delta_i$ with perturbation magnitude proportional to $\|v_i\|_2$. For structure-based attacks such as permutation, $\phi_i(V)=v_{\sigma(i)}$ for some derangement $\sigma$, and the resulting discrepancy depends on the pairwise token difference $\|v_{\sigma(i)}-v_i\|_2$, which is bounded by the overall token magnitude, e.g.,
$\|v_{\sigma(i)}-v_i\|_2 \le 2\|V\|_F$.

\noindent\textbf{Proof.} Let $G(V)=VV^\top$ denote the pairwise similarity matrix of vision tokens. Suppose a selected token $v_i$ is perturbed to
\begin{equation}
\tilde v_i = \phi_i(V)=v_i+\Delta_i,
\end{equation}
where
\begin{equation}
\Delta_i := \phi_i(V)-v_i
\end{equation}
denotes the effective perturbation induced by the transformation operator $\phi_i$. All other tokens remain unchanged.

Then, for any $j\neq i$, we have
\begin{equation}
\tilde G_{ij}-G_{ij}
=
\tilde v_i^\top v_j - v_i^\top v_j
=
(v_i+\Delta_i)^\top v_j - v_i^\top v_j
=
\Delta_i^\top v_j,
\end{equation}
and similarly
\begin{equation}
\tilde G_{ji}-G_{ji}=\Delta_i^\top v_j.
\end{equation}
For the diagonal entry,
\begin{equation}
\tilde G_{ii}-G_{ii}
=
\|\tilde v_i\|_2^2-\|v_i\|_2^2
=
\|v_i+\Delta_i\|_2^2-\|v_i\|_2^2
=
2v_i^\top\Delta_i+\|\Delta_i\|_2^2.
\end{equation}
All other entries remain unchanged. Therefore,
\begin{equation}
\|\tilde G-G\|_F^2
=
2\sum_{j\ne i}(\Delta_i^\top v_j)^2
+
\left(2v_i^\top\Delta_i+\|\Delta_i\|_2^2\right)^2.
\end{equation}

Using the Cauchy--Schwarz inequality,
\begin{equation}
(\Delta_i^\top v_j)^2 \le \|\Delta_i\|_2^2\|v_j\|_2^2,
\end{equation}
which implies
\begin{equation}
\|\tilde G-G\|_F^2
\le
2\|\Delta_i\|_2^2\sum_{j\ne i}\|v_j\|_2^2
+
\left(2v_i^\top\Delta_i+\|\Delta_i\|_2^2\right)^2.
\end{equation}

This shows that the change in the similarity matrix is governed by the effective perturbation magnitude $\|\Delta_i\|_2$. Under the assumption that the induced discrepancy is controlled by the token magnitude or, more generally, bounded by the overall token energy, tokens associated with larger effective perturbations tend to induce larger changes in $VV^\top$.

Since the self-attention operator is a smooth function of $VV^\top$, larger perturbations to $VV^\top$ generally lead to larger deviations in the resulting attention patterns. Consequently, when maximizing the self-attention disruption objective $\mathcal{L}_{\mathrm{SA}}$, the optimization tends to favor tokens whose transformation induces larger effective discrepancies in the similarity structure.

\subsection{Adaptive $\lambda$ Calibration}
\label{appendixsec:optimization}

The relative scales of $\mathcal{L}_{\mathrm{SA}}$ and $\mathcal{L}_{\mathrm{norm}}$ vary across inputs due to differences in vision token distributions. A fixed $\lambda$ cannot maintain a consistent balance between the two terms. We therefore calibrate $\lambda$ adaptively. Given the uniform initialization $w_i^{(0)} = \frac{k}{L_v}$, we perform a single forward-backward pass to estimate the average gradient magnitudes:
\begin{equation}
\bar{g}_{\mathrm{SA}} =
\mathrm{mean}\!\left(
\left|
\nabla_{{w}} \mathcal{L}_{\mathrm{SA}}
\right|
\right),
\quad
\bar{g}_{\mathrm{norm}} =
\mathrm{mean}\!\left(
\left|
\nabla_{{w}} \mathcal{L}_{\mathrm{norm}}
\right|
\right).
\end{equation}
We then set
\begin{equation}
\lambda = \alpha \cdot \frac{\bar{g}_{\mathrm{SA}}}{\bar{g}_{\mathrm{norm}}},
\end{equation}
with $\alpha = 0.5$. 
By setting $\lambda$ proportional to $\bar{g}_{\mathrm{SA}} / \bar{g}_{\mathrm{norm}}$, the gradient magnitude of the regularization term is adjusted to be comparable to that of the self-attention disruption term at the starting point. The coefficient $\alpha$ further controls the relative strength of the regularization, allowing $\mathcal{L}_{\mathrm{SA}}$ to remain the primary objective while preventing the optimization from being overly biased toward high-norm tokens.

\end{document}